\documentclass{ifacconf}

\usepackage{graphicx}      % include this line if your document contains figures
\usepackage{natbib} % required for bibliography

\usepackage{url}
\usepackage{amsmath}
\usepackage{amssymb}
\usepackage{booktabs}
\usepackage{mathtools}
\usepackage[frozencache]{minted}
\usepackage{listings}
\newcommand{\rBr}[1]{\left(#1\right)}

\newcommand{\Real}{\mathbb{R}}
\newcommand{\tens}[1]{\boldsymbol{\mathcal{#1}}} % Tensor calligraphic
\newcommand{\mat}[1]{\boldsymbol{#1}} % Matrix bold
\newcommand{\vect}[1]{\boldsymbol{#1}} % Vectors bold
\newcommand{\inRe}[1]{\in\Real^{#1}} % \in\Real{}
\begin{document}

\begin{frontmatter}
\title{Tensor Network Kernel Machines: A JAX Framework for Machine Learning and Nonlinear System Identification} 

\thanks[footnoteinfo]{This work is supported by the
Dutch Research Council (NWO).}

\author[First]{Albert Saiapin} 
\author[First]{Kim Batselier} 

\address[First]{Delft Center for Systems and Control, TU Delft, Netherlands (e-mail: \{a.saiapin, k.batselier\}@tudelft.nl)}

\begin{abstract}
Developing nonlinear models that are both expressive and computationally
efficient remains a challenge in machine learning and nonlinear system
identification. Tensor network kernel machines (TNKM) address this
challenge by combining nonlinear feature representations with compact
low-rank tensor-network parameterizations. However, practical and
extensible software frameworks for developing TNKM models remain limited.
In this work, we introduce \emph{tnkm}, an open-source Python library for constructing and training TNKM models using \emph{JAX}. The library
provides a unified interface for combining different feature maps,
tensor-network architectures, and optimization strategies, including
alternating least squares and gradient-based methods.
We demonstrate the capabilities of \emph{tnkm} on nonlinear benchmark
problems, showing that the implemented models achieve competitive
prediction accuracy while retaining compact parameterizations and
efficient training. The proposed framework facilitates reproducible
development and application of tensor-network-based learning methods.
\end{abstract}

%\begin{keyword}
%Tensor Networks; Kernel Methods; System Identification; Machine Learning;
%\end{keyword}

\end{frontmatter}
%===============================================================================

\section{Introduction}\label{sec:intro}
Nonlinear function approximation is a fundamental problem in both machine learning (ML) and nonlinear system identification, where the objective is to learn accurate input--output relationships directly from data. Such models underpin a wide range of applications, including computer vision, natural language processing, robotics, autonomous systems, process control, and scientific computing~\citep{ML_Bishop_2006,dl_2016_goodfellow,nsi_1999_ljung}. As these applications continue to grow in complexity and scale, developing nonlinear models that combine high predictive accuracy with computational efficiency while remaining scalable to high-dimensional problems remains an important research challenge.

Over the past decades, numerous approaches have been developed for nonlinear modeling, including kernel methods, Gaussian processes, and deep neural networks, each offering different trade-offs between predictive performance, computational complexity, and interpretability~\citep{kernel_2001_smola,gp_2006_rasmussen,dl_2015_lecun}. 
Many of these methods, however, face scalability challenges arising from either rapidly growing model complexity or increasing computational costs as the number of model parameters or training samples grows. Tensor network kernel machines (TNKM)~\citep{TD_FF_Wesel_2021,lrtdnsi_2022_batselier} address this challenge by combining explicit nonlinear feature representations with compact low-rank tensor-network parameterizations of the model coefficients. By exploiting tensor decompositions such as the canonical polyadic (CP) and tensor-train (TT) formats, TNKM can represent high-dimensional nonlinear models using substantially fewer parameters than their full tensor counterparts. These properties have made TNKM an effective approach for nonlinear regression, classification, and system identification~\citep{EM_Novikov_2018,la_ttkm_saiapin_2025,la_tnkm_saiapin_2026}.

Despite the growing interest in tensor-network-based machine learning~\citep{TN1_Cichocki_2016,tn_app_2017_cichocki}, software frameworks for
these models remain diverse and target different application scenarios.
Existing libraries provide valuable tools for tensor-network optimization,
ML applications~\citep{tensorly_2019_kossaifi}, and quantum
systems simulations~\citep{tenpy_2024_Hauschild,quimb_2018_gray}.
More closely related to the present work, \emph{tn4ml}~\citep{tn4ml_2025_puljak}
provides a configurable framework for training tensor-network-based
ML models. While these frameworks provide important foundations for tensor-network-based learning, this work focuses specifically on the tensor network kernel machine paradigm, providing a unified framework tailored to machine learning and nonlinear system identification applications.

In this work, we introduce \emph{tnkm}~\footnote{https://github.com/AlbMLpy/tnkm}, an open-source Python library
for developing and applying tensor network kernel machines. The library
provides a modular framework that separates the main components of TNKM
models, including nonlinear feature maps, tensor-network
parameterizations, and optimization algorithms. This design enables
systematic comparison of different TNKM configurations and facilitates
the extension of the framework with new representations and training
strategies. The main contributions of this work are:
\begin{itemize}
    \item We introduce a unified \emph{JAX}-based framework for TNKM models with minimal external dependencies, supporting multiple tensor-network parameterizations, including canonical polyadic (CP) and tensor-train (TT) representations, within a common interface.
    \item We provide scalable training implementations combining
    structure-aware optimization through alternating least squares (ALS)
    with general gradient-based optimization methods, enabling
    application to both regression and broader learning tasks.
    \item We develop a modular software architecture with reusable interfaces for feature maps, model classes, and optimization routines, complemented by practical utilities for nonlinear system identification, including data preprocessing, simulation, and evaluation.
    \item We present a comprehensive experimental evaluation of the proposed framework, including predictive-performance benchmarks and optimization studies, demonstrating the effectiveness and practical applicability of TNKM models.
\end{itemize}
The remainder of this paper is organized as follows. Section~\ref{sec:method}
introduces the mathematical formulation of tensor network kernel machines.
Section~\ref{sec:software} describes the design and implementation of the
\emph{tnkm} library. Section~\ref{sec:experiments} presents experimental results on nonlinear benchmark problems and compares the supported optimization methods. Finally, Sections~\ref{sec:discussion} and~\ref{sec:conclude} discuss the implications and limitations of the proposed framework, as well as opportunities for its future development.

\section{Methodology}\label{sec:method}
\begin{figure}[!t]
\begin{center}
    \includegraphics[width=\linewidth]
    {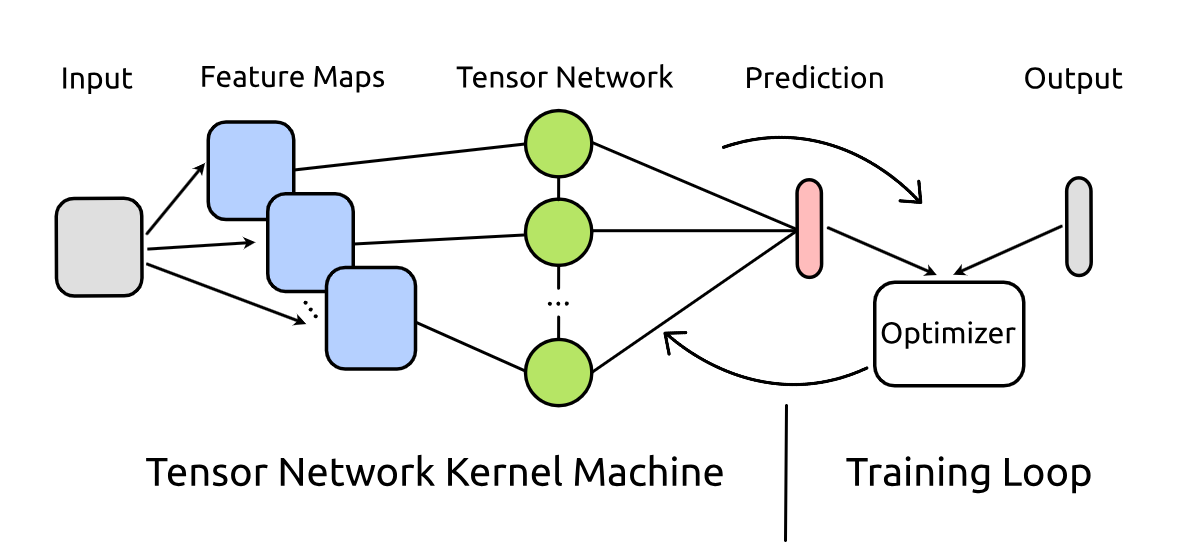}
    \caption{Overview of the tensor network kernel machine (TNKM) framework. A TNKM model combines a feature map, a low-rank tensor-network parameterization, and an optimization algorithm within one framework.}
    \label{fig:tnkm_lib_diagram}
\end{center}
\end{figure}

Tensor network kernel machines combine nonlinear feature
representations with low-rank tensor parameterizations to construct
expressive yet computationally tractable nonlinear models. To reflect
this formulation, the \emph{tnkm} library adopts a modular architecture
that separates feature maps, tensor-network parameterizations, and
optimization algorithms into independent components. An overview of the
framework and the relationships between these components is shown in
Figure~\ref{fig:tnkm_lib_diagram}. The following subsections describe
each component and its mathematical formulation in detail.

\subsection{Tensor network kernel machines}\label{sec:tnkm}
Consider a nonlinear regression problem with input
$\vect{x}\inRe{D}$ and target $y\inRe{}$. A common
approach is to represent the nonlinear mapping using a linear model in a
high-dimensional feature space~\citep{kernel_2001_smola,ML_Bishop_2006},
\begin{equation}\label{eq:model_form}
    f(\vect{x}) = \vect{\phi}(\vect{x})^\top\vect{w},
\end{equation}
where $\vect{\phi}(\vect{x})$ denotes a nonlinear feature map
and $\vect{w}$ is the vector of trainable model parameters. The choice of the feature map $\vect{\phi}$ determines the expressive power of the resulting model and enables approximation of a wide class of nonlinear functions.
For multivariate inputs, however, the dimensionality of $\vect{\phi}(\vect{x})$ typically grows exponentially with the number of input variables, making direct optimization of $\vect{w}$ computationally prohibitive. Tensor network kernel machines (TNKM) address this challenge by exploiting two key ideas.

First, the feature map is assumed to admit a tensor-product representation,
\begin{equation}\label{tensor_product_features}
\vect{\phi}(\vect{x}) \coloneq \vect{\phi}^{(D)}(x_D) \otimes \dots \otimes \vect{\phi}^{(1)}(x_1),
\end{equation}
where $\vect{\phi}^{(d)}: \mathbb{R} \rightarrow \mathbb{R}^{I_d}$
denotes a one-dimensional feature map acting on the $d$-th input variable. Such tensor-product feature maps naturally arise from product kernels~\citep{VFF_Hensman_2017,GPRR_Solin_2019}. 

Second, instead of storing the parameter vector $\vect{w}$ explicitly, the model parameters are represented using a low-rank tensor-network decomposition. This substantially reduces the number of trainable parameters while preserving the multilinear structure required for
efficient optimization and inference~\citep{TD_FF_Wesel_2021, QTNM_Wesel_2024, lrtdnsi_2022_batselier}. Rather than storing $\prod_{d=1}^D I_d$ parameters explicitly, low-rank tensor decompositions reduce the parameter complexity to $\mathcal{O}(DIR)$ for CP decompositions and $\mathcal{O}(DIR^2)$ for TT decompositions, where $I=\max(I_1, \dots, I_D)$. For example, if each feature map contains $I_d=20$ features, $D=10$, and $R=10$, the full parameter vector $\vect{w}$ contains $20^{10} \approx 10^{13}$ coefficients, making direct optimization infeasible, whereas a low-rank CP representation requires only about $2 \times 10^3$ parameters.

With the feature map $\vect{\phi}$ fixed, training reduces to estimating the low-rank tensor parameters from the training data
$\mathcal{D} \coloneq \{\rBr{\vect{x}_i, y_i} \mid \vect{x}_i \inRe{D},
\, y_i \inRe{},\; i=1,\dots,N\}$. The feature map specifies the nonlinear
embedding of the inputs, while the tensor-network parameterization
determines the complexity of the resulting TNKM model.

\subsection{Feature maps}
As discussed in the previous section, the feature map
$\vect{\phi}$ determines the nonlinear representation of the input. Its
choice therefore directly influences the class of nonlinear functions
that can be represented by the model. TNKM constructs multivariate
feature representations from one-dimensional feature maps according to
\eqref{tensor_product_features}. This formulation allows different
feature maps to be combined with the same tensor-network
parameterization and optimization algorithms, resulting in a flexible
and modular modeling framework.

The \emph{tnkm} library supports several classes of feature maps suited to different modeling scenarios. In this paper, we focus on three representative families: polynomial-based, Fourier-based, and memory-based (Volterra) features. Note that some of these feature maps assume inputs normalized to the interval $[0,1]$; accordingly, appropriate preprocessing should be applied prior to training.

\subsubsection{\textbf{Polynomial features.}}
Polynomial features~\citep{poly_tt_2016} construct nonlinear
representations by embedding each input variable into a polynomial
basis. The local feature map is defined as
\begin{equation}\label{eq:poly}
    \vect{\phi}^{(d)}(x) =\left[1, x, x^2, \dots, x^{I_d-1}\right]^\top,
\end{equation}
with $x \inRe{}$.
These features are particularly suitable for problems whose
nonlinearities can be well approximated by low-degree polynomial
interactions. Increasing the embedding dimension $I_d$ increases the
expressive power of the model at the cost of a larger feature space.
More generally, by the Stone--Weierstrass theorem~\citep{swt_branges_1959}, polynomial bases are capable of approximating any continuous function on a compact domain to arbitrary accuracy.

\subsubsection{\textbf{Fourier features.}}
Fourier features~\citep{QTNM_Wesel_2024} are widely used in kernel methods, as they correspond to the eigenfunctions of $D$-dimensional stationary product kernels with respect to the Lebesgue measure~\citep{VFF_Hensman_2017}. For an input value $x\inRe{}$, the local feature map is defined as
\begin{equation}\label{ff_map}
    \vect{\phi}^{(d)}(x) = c_d \left[e^{-\frac{2\pi j k}{L} x}\right]_{k = 0}^{I_d - 1},
\end{equation}
where $j$ denotes the imaginary unit, $c_d = e^{2\pi j x \frac{2+I_d}{2L}}$ is a centering factor, and $L$ is the period hyperparameter defining the function class. These features are especially effective at approximating smooth, stationary functions and nonlinearities exhibiting periodic or oscillatory behavior. The embedding dimension $I_d$ determines the number of Fourier basis functions used to represent the input, with larger values enabling more complex functions to be approximated. The period parameter $L$ controls the range of frequencies captured by the basis and should be chosen according to the expected scale of variation in the data.

\subsubsection{\textbf{Volterra features.}}
Volterra features~\citep{tnkf_batselier_2017} are designed for nonlinear dynamical systems and sequence modeling, where the current output $y_t$ depends not only on the current input $u_t$ but also on previous inputs $u_{t-1}, u_{t-2}, \dots$. Instead of embedding a single input value, the feature map operates on a finite window of the input sequence.

Given an input sequence $\vect{u}=[u_1,\ldots,u_T]^\top\inRe{T}$, the local feature map at time instant $t$ is defined as
\begin{equation}\label{volt_fmap}
    \vect{\phi}_t(\vect{u}) = \left[1, u_t, u_{t-1}, \dots, u_{t-I+1}\right]^\top,
\end{equation}
where $I$ denotes the embedding dimension (or memory length).
Higher-order Volterra interactions are obtained by constructing the tensor-product feature representation
\begin{equation}
    \underbrace{\vect{\phi}_t(\vect{u}) \otimes \cdots \otimes \vect{\phi}_t(\vect{u})}_{D\ \mathrm{times}},
\end{equation}
where the tensor order $D$ specifies the maximum order of the Volterra expansion.

\subsection{Low-rank tensor parameterizations}
As discussed in Section~\ref{sec:tnkm}, the main computational challenge
of model~\eqref{eq:model_form} is the exponentially large parameter
vector $\vect{w}$. Tensor networks (TNs)~\citep{TN1_Cichocki_2016}
address this problem by representing high-dimensional tensors as
networks of smaller low-rank tensors (cores) connected through latent
indices. This representation dramatically reduces the number of
trainable parameters while preserving the multilinear structure of the
model.

To enable such a representation, the parameter vector $\vect{w}$ is interpreted as a $D$-th order tensor $\tens{W}\inRe{I_1\times I_2\times\cdots\times I_D}$ according to
\begin{equation}
    w_i \coloneq W_{i_1i_2\dots i_D},
\end{equation}
where
\[
i \coloneq i_1+\sum_{d=2}^{D}i_d\prod_{j=1}^{d-1}I_j.
\]

The \emph{tnkm} library supports two widely used tensor-network parameterizations: the canonical polyadic (CP) decomposition and the tensor-train (TT) decomposition. These representations provide different trade-offs between compactness, expressiveness, and computational cost, allowing users to select the tensor-network backbone that best matches the requirements of a given application through a unified interface.

\subsubsection{\textbf{CP decomposition.}}
CP decomposition~\citep{TD_Kolda_2009} represents the model parameters as a sum of rank-one components,
\begin{equation}
    \vect{w}(\vect{v}) = \sum_{r=1}^{R} \vect{v}^{(D)}_r\otimes\cdots\otimes\vect{v}^{(1)}_r ,
\end{equation}
where $R$ is the CP rank and $\vect{v}$ collects all trainable
parameters of the decomposition. Instead of storing $\prod_{d=1}^{D}I_d$ coefficients explicitly, the CP representation requires only $R\sum_{d=1}^{D}I_d$ trainable parameters, providing a compact low-rank parameterization. The rank $R$ controls the trade-off between model complexity and expressive power.

\subsubsection{\textbf{TT decomposition}}
TT decomposition~\citep{TT_Oseledets_2011}
parameterizes the model weights as a sequence of interconnected tensor cores,
\begin{equation}
    w_{i} = \mat{V}^{(1)}(i_1) \mat{V}^{(2)}(i_2) \cdots \mat{V}^{(D)}(i_D),
\end{equation}
where $\mat{V}^{(d)}(i_d)\inRe{R_{d-1}\times R_d}$ denotes the
$i_d$-th matrix slice of the $d$-th TT core, and $R_0=R_D=1$. TT-ranks determine the expressive power of the representation while controlling its complexity. The resulting number of trainable parameters is $\sum_{d=1}^{D}I_dR_{d-1}R_d$, which grows linearly with the number of dimensions for fixed TT ranks.

\subsection{Optimization methods}
The objective of training TNKM is to estimate the trainable parameters
$\vect{v}$ from a dataset $\mathcal{D}$ by minimizing a regularized
empirical loss,
\begin{equation}
    \min_{\vect{v}}\sum_{n=1}^{N} \ell\!\left(y_n, f(\vect{x}_n;\vect{v})\right) + \mathcal{R}(\vect{w}(\vect{v})),
\end{equation}
where $\ell(\cdot,\cdot)$ denotes the loss function and
$\mathcal{R}(\vect{w}(\vect{v}))$ is an optional regularization term. Two regularization strategies can be considered depending on whether the penalty is applied to the represented model coefficients or to the tensor-network parameters. Regularization of $\vect{w}$ directly controls the complexity of the resulting nonlinear model, while regularization of $\vect{v}$ constrains the low-rank representation and provides a computationally cheaper alternative.

The \emph{tnkm} library provides two complementary optimization strategies for training TNKM models: alternating least squares (ALS) and gradient-based optimization. The choice of optimizer therefore depends on the selected tensor-network representation and the desired balance between computational efficiency and modeling flexibility.

\subsubsection{\textbf{Alternating least squares.}}
Alternating least squares (ALS) exploits the multilinear structure of
tensor-network parameterizations by optimizing one tensor core at a time
while keeping all remaining cores fixed~\citep{td_als_2009_comon, tt_als_2012_holtz}. Each subproblem reduces to a linear least-squares problem with a closed-form solution, resulting in deterministic updates that typically converge in only a few iterations. ALS is particularly effective for CP and TT parameterizations with quadratic loss functions, providing fast optimization with minimal hyperparameter tuning.

\subsubsection{\textbf{Gradient-based optimization.}}
As an alternative, the \emph{tnkm} library supports gradient-based optimization through automatic differentiation and the \emph{Optax} optimization
library~\citep{jax_deepmind_2020}. This approach is applicable to arbitrary differentiable models and loss functions, making it considerably more flexible than ALS. The library provides several first-order optimization algorithms, including SGD~\citep{sgd_2010_bottou}, RMSProp~\citep{rnn_2014_graves}, Adam~\citep{adam_kingma_2017}, among others, allowing users to select the optimizer best suited to their application.

ALS and gradient-based optimization provide complementary trade-offs between computational efficiency and modeling flexibility. ALS is highly efficient for multilinear least-squares problems and is the recommended choice whenever applicable, whereas gradient-based methods naturally extend to arbitrary differentiable models and are therefore applicable to a broader range of learning problems.

\section{Software Design}\label{sec:software}
\subsection{Library architecture}

\begin{figure}[!t]
\begin{center}
    \includegraphics[width=\linewidth]
    {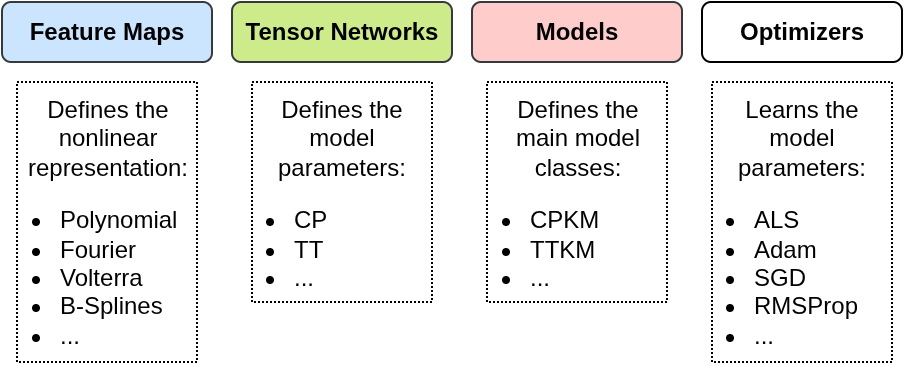}
    \caption{Architecture of the \emph{tnkm} library, showing the main components used to construct and train tensor network kernel machines.
    }
    \label{fig:tnkm_design}
\end{center}
\end{figure}

The implementation of the \emph{tnkm} library follows the modular formulation introduced in the previous section. As illustrated in Figure~\ref{fig:tnkm_design}, a TNKM model is assembled by combining four main components: (i) a feature map that defines the nonlinear representation of the inputs, (ii) a tensor-network backbone that specifies the low-rank parameterization of the model coefficients and (iii) is implemented through the corresponding model classes (e.g., \emph{CPKM} and \emph{TTKM}), and (iv) an optimization algorithm used to estimate the trainable parameters. This separation allows different feature maps, tensor-network representations, and optimization algorithms to be combined within a unified interface, enabling flexible model development and efficient experimentation.

In addition to these core components, the library provides auxiliary modules for common tasks such as data preprocessing, data loading, simulation, and model evaluation. These utilities support the end-to-end workflow but are not central to the TNKM formulation and are therefore not discussed in detail here. Their complete description and usage examples are provided in the library documentation.

The computational backend of \emph{tnkm} is based on the
\emph{JAX} library~\citep{jax_2018_github}, which provides automatic
differentiation and just-in-time compilation for efficient numerical
computations. Gradient-based optimization is implemented using
\emph{Optax}~\citep{jax_deepmind_2020}, which provides a range of
\emph{JAX}-compatible first-order optimization algorithms.

\subsection{API and usage}
Listing~\ref{listing:example} demonstrates the typical workflow for
constructing and training a TNKM model using the \emph{tnkm} library.
This example uses a CP-based model with polynomial feature maps and
alternating least squares optimization.

\begin{listing}[!t]
\begin{minted}[
bgcolor=gray!5,
]{python}
import jax.numpy as jnp

from tnkm.models import CPKM
from tnkm.optim import train_als
from tnkm.features import (
    PolyFeature,
    ProductFeatures
)

# Training data
X_train = jnp.ones((32, 2))
y_train = jnp.ones((32,))

# Construct feature map
feature_map = ProductFeatures(
    [
        PolyFeature(4, k_col=i) 
        for i in range(X_train.shape[1])
    ]
)

# Initialize model
model = CPKM(
    feature_map, 
    rank=2,
    seed=0
)

# Train model
train_als(
    model, 
    X_train, 
    y_train, 
    n_epoch=3, 
    gamma_w=1e-3, 
    beta_e=1.0
)

# Prediction
y_pred = model.predict(X_train)
\end{minted}
\vspace{1em}
\caption{Example workflow for defining and training a tensor network kernel machine.}
\label{listing:example}
\end{listing}

The feature map is constructed as a product of one-dimensional
polynomial embeddings, while the CP rank controls the complexity of the
tensor-network parameterization. The trained model can subsequently be
used for prediction. This example illustrates the correspondence between
the mathematical formulation of TNKM and its software implementation.
The modular design allows individual components to be exchanged without
changing the overall workflow.

\section{Experiments and Applications}\label{sec:experiments}
This section demonstrates the capabilities of the \emph{tnkm} library from two complementary perspectives. First, we compare alternating least squares and gradient-based optimization on a regression task to study their convergence behavior, predictive performance, and practical trade-offs. Second, we evaluate the proposed framework on established nonlinear system identification benchmarks by comparing its performance against representative methods using recursive simulation.

All experiments were conducted on a Dell Latitude 7440 laptop equipped with a 13th Gen Intel Core i7-1365U processor and 16\,GB of RAM. The complete source code, together with all implementation details and the data required to reproduce the reported results, is publicly available on GitHub under the \emph{experiments/} directory.\footnote{https://github.com/AlbMLpy/tnkm}

\subsection{Optimization methods}
The \emph{tnkm} library supports multiple optimization strategies through a unified API,
including alternating least squares (ALS) and gradient-based
optimization. This flexibility allows users to select the optimization
strategy that best matches their application and computational
requirements. In this section, we compare the optimization behavior of
ALS and Adam and discuss the practical trade-offs between the two
approaches.

The comparison is performed on the Airfoil Self-Noise regression
dataset~\citep{airfoil_Brooks_1989}, a widely used benchmark for
supervised learning. The data are randomly split into training and validation
subsets using a 90/10 ratio. All experiments use the same TNKM configuration consisting of squared exponential (Fourier) features with $I_d = 20$ frequencies, a CP-based kernel machine of rank $R=10$, $\mathcal{R}=\ell_2$ regularization of the tensor cores $\vect{v}$, and fixed random seeds.

Gradient-based optimization is performed using the Adam
optimizer~\citep{adam_kingma_2017} with both mini-batch and full-batch
updates. In contrast, ALS optimizes the model by alternating exact
updates of the tensor-network cores. As a reference, we additionally
report the validation error of kernel ridge regression (KRR)~\citep{lssvm_2002_suykens}, which
corresponds to the unconstrained solution of the underlying regression
problem~\citep{TD_FF_Wesel_2021}.

\begin{figure}[!t]
\begin{center}
    \includegraphics[width=\linewidth]
    {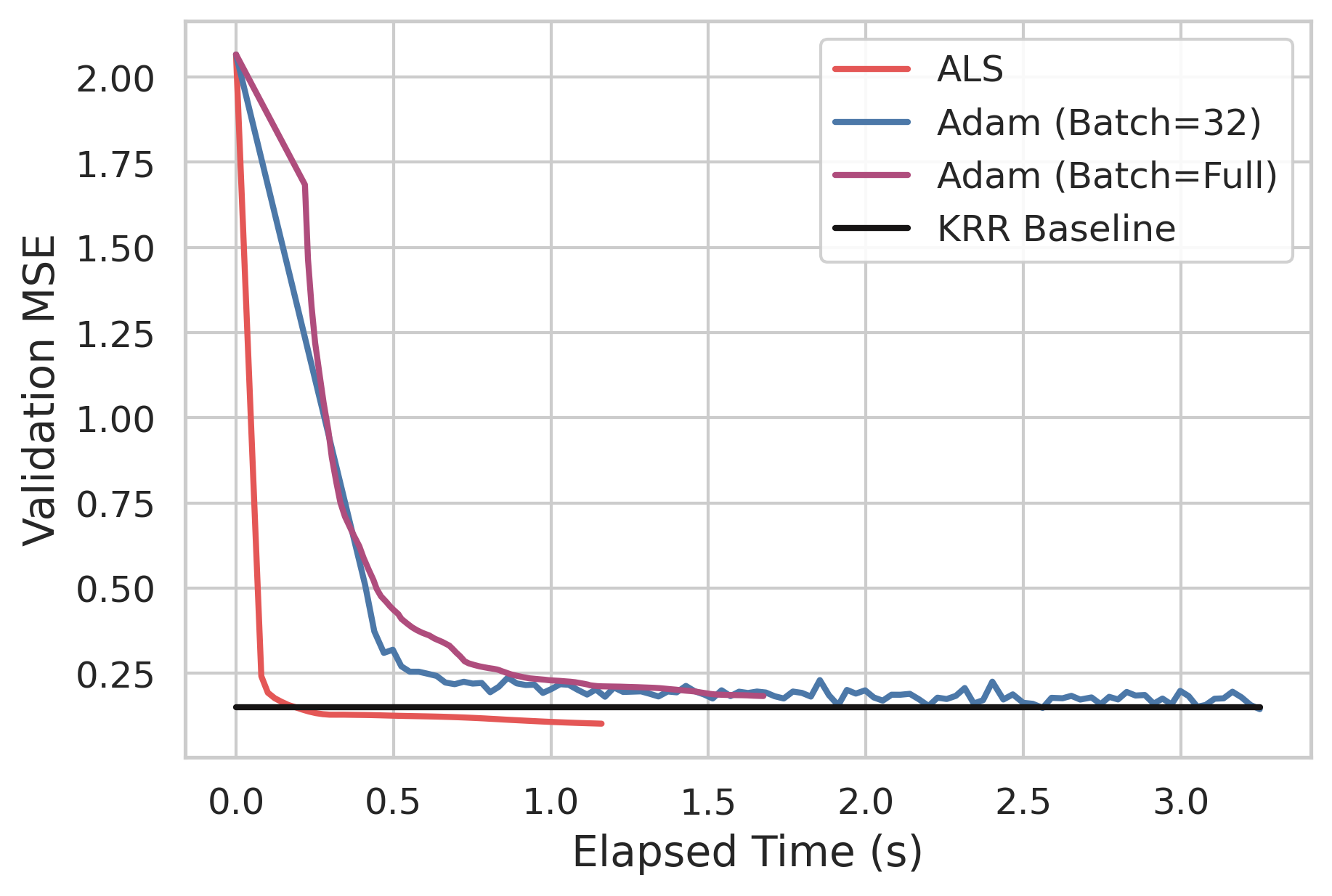}
    \caption{Validation MSE versus elapsed training time for ALS and Adam on the Airfoil Self-Noise dataset. The horizontal dashed line indicates the validation MSE achieved by kernel ridge regression (KRR).}
    \label{fig:opt}
\end{center}
\end{figure}

Table~\ref{table:optim} summarizes the final validation MSE and training
time, while Figure~\ref{fig:opt} shows the corresponding optimization
trajectories as a function of elapsed training time. The results show that ALS converges
rapidly to a solution with validation error close to the KRR reference,
while requiring only a fraction of the training time of Adam. Moreover,
its convergence is stable throughout the optimization process. Adam also
approaches the KRR baseline but generally requires substantially more
optimization steps and training time. As expected, full-batch training
converges faster than the mini-batch variant due to the reduced
stochasticity of the gradient estimates. Nevertheless, gradient-based
optimization offers greater flexibility, supporting arbitrary
differentiable objectives and model formulations beyond the least-squares
setting for which ALS is specifically designed.

\begin{table}
\begin{center}
\caption{Comparison of optimization methods on the Airfoil Self-Noise dataset. KRR is included as a reference solution.}
\label{table:optim}
\begin{tabular}{||l||c|c|c||}
\toprule
Method & Validation MSE & Time(s) & Epochs \\
\midrule
ALS & 0.102 & 1.161 & 50 \\
Adam (Batch=32) & 0.145 & 3.251 & 100 \\
Adam (Batch=Full) & 0.183 & 1.675 & 100 \\
\midrule
KRR Baseline & 0.15 & N/A & N/A \\
\bottomrule
\end{tabular}
\end{center}
\end{table}

%\begin{table} \begin{center} \end{center} \end{table}

\subsection{Nonlinear system identification benchmarks}
The nonlinear system identification study comprises three benchmark
problems: Silverbox~\citep{silverbox_Schoukens_2013}, Coupled Electric Drives~\citep{ced_Schoukens_2017}, and Cascaded
Tanks~\citep{tanks_Schoukens_2017}. These benchmarks cover a range of nonlinear dynamical systems with different characteristics and data sizes, providing a diverse evaluation of the proposed framework.

For all benchmarks, the input and output variables are scaled using min--max normalization. Given a variable $x$, the normalized value is computed as
\begin{equation*}
    \tilde{x}=\frac{x-x_{\min}}{x_{\max}-x_{\min}},
\end{equation*}
where $x_{\min}$ and $x_{\max}$ are computed exclusively from the
training data. 
All TNKM models are formulated using a NARX representation:
\begin{equation}
    y_n = f(\{y_{n-i}\}_{i \in \mathcal{I}_y}, \{u_{n-j}\}_{j \in \mathcal{I}_u}),
\end{equation}
where $\mathcal{I}_y$ and $\mathcal{I}_u$ denote the corresponding output and input lag sets. The lag orders are selected heuristically based on partial autocorrelation analysis. Model hyperparameters are selected using cross-validation on the training (validation) data, while the final models are evaluated on an unseen test set. Performance is assessed using recursive (free-run) simulation, where an initial measured output history is used to initialize the NARX regressor and the predicted outputs are subsequently fed back into the model. Predictive accuracy is quantified using the root mean squared error (RMSE), where lower values indicate better performance.

To evaluate the performance of TNKM, we compare its results with
representative nonlinear system identification methods, including
dynoNet~\citep{dynoNet_2021_forgione}, SUBNET~\citep{subnet_2021_Beintema}, CT-SUBNET~\citep{ct_subnet_2023_Beintema}, Normalized Gray-box SSNN~\citep{ngbssnn_2026_Forgione}, WP-LFR~\citep{WP-LFR_2026_Drenth}, NL-LFR~\citep{nf_lfr_2025_floren}, and other established approaches. The baseline results are obtained from the official Nonlinear Benchmark website~\footnote{https://www.nonlinearbenchmark.org/benchmarks}, which provides a standardized collection of benchmark results for nonlinear system identification. All the tables report RMSE on the corresponding test sequences (e.g., RMSE(1) denotes the metric on Test 1) and an approximate white-box/black-box (W/B) interpretability score, where 1 denotes a fully white-box model and 10 a fully black-box model.

\subsubsection{\textbf{Coupled Electric Drives.}}

\begin{table}
\begin{center}
\caption{Comparison of identification methods on the Coupled Electric Drives benchmark.}
\label{table:ced}
\begin{tabular}{||l||c|c|c|c||}
\toprule
Method & RMSE(1) & RMSE(2) & Time & W/B \\
\midrule
dynoNet & 0.062 & 0.047 & 1-60 sec & 10 \\
GPNARX & 0.087 & 0.070 & 1-60 min & 10 \\
RNN & 0.114 & 0.138 & 1-60 min & 10 \\
CT-SUBNET & 0.115 & 0.074 & 1-60 min & 9 \\
MLPNARX & 0.124 & 0.194 & 1-60 sec & 10 \\
LSTM & 0.139 & 0.111 & 1-60 min & 10 \\
\midrule
TNKM & 0.085 & 0.077 & $<$ 0.1 sec & 10 \\
\bottomrule
\end{tabular}
\end{center}
\end{table}

%\begin{table} \begin{center} \end{center} \end{table}

\begin{figure}[!t]
\begin{center}
    \includegraphics[width=\linewidth]
    {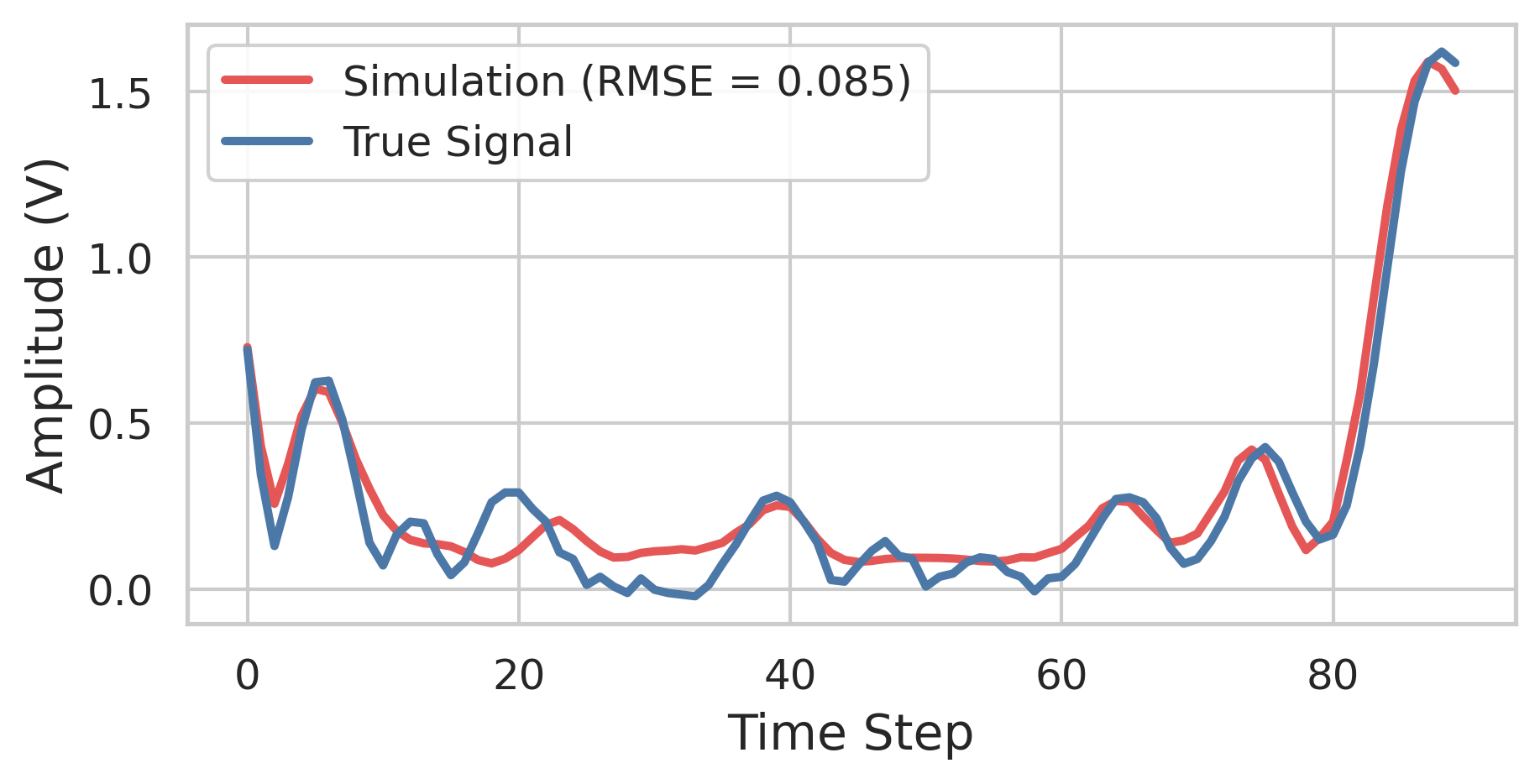}
    \caption{Simulation results on the Coupled Electric Drives benchmark (Test 1). The measured output is compared with the output predicted by the identified TNKM model.}
    \label{fig:ced_test_1}
\end{center}
\end{figure}

The Coupled Electric Drives benchmark describes a nonlinear mechanical system consisting of two electric motors driving a pulley through a flexible belt. The pulley is connected to a spring, introducing a lightly damped dynamic mode. In this study, we focus on the speed control setup, where the angular velocity of the pulley is used as the system output.

This benchmark provides two training and two test datasets, containing 400 and 100 samples per sequence, respectively. The first 10 measured output samples of each test sequence are used to initialize the NARX regressor. The input and output lag sets are defined as $\mathcal{I}_u = \{0,1,2,3,4,6,8,10\}$ and $\mathcal{I}_y = \{1,2,3,4,6,8,10\}$, respectively. The two training datasets are combined to form the final training set. The selected TNKM configuration uses a CP kernel machine (CPKM) architecture with rank $R=15$ and polynomial feature maps of degree $I_d=3$.

Table~\ref{table:ced} summarizes the performance of TNKM and the
considered baseline methods. Figure~\ref{fig:ced_test_1} shows the
recursive simulation results for the first test sequence; the second test
sequence exhibits similar behavior. The results indicate that TNKM achieves predictive performance comparable to the best-performing methods, matching GPNARX and only slightly underperforming dynoNet in terms of RMSE. At the same time, TNKM requires less than one second for training, whereas the reported training times for the competing methods range from several seconds to minutes.

\subsubsection{\textbf{Cascaded Tanks.}}

The Cascaded Tanks benchmark describes a nonlinear fluid level control system consisting of two tanks, a water reservoir, and a pump. The pump voltage is used as the system input, while the water level of the lower tank is considered as the system output. Water is pumped into the upper tank, flows through an opening into the lower tank, and finally returns to the reservoir. The benchmark contains both smooth nonlinearities arising from the fluid dynamics and a hard saturation nonlinearity caused by overflow of the upper tank. 

The benchmark provides training and test datasets containing 1024 samples per sequence. The NARX regressor is initialized using the first 50 measured output samples of each test sequence. The input and output lag sets are defined as $\mathcal{I}_u = \mathcal{I}_y = \{1,2,3,4,8,12,16,32\}$. The selected TNKM configuration uses a CPKM backbone with rank $R=13$ and polynomial feature maps of degree $I_d=2$.

\begin{figure}[!t]
\begin{center}
    \includegraphics[width=\linewidth]{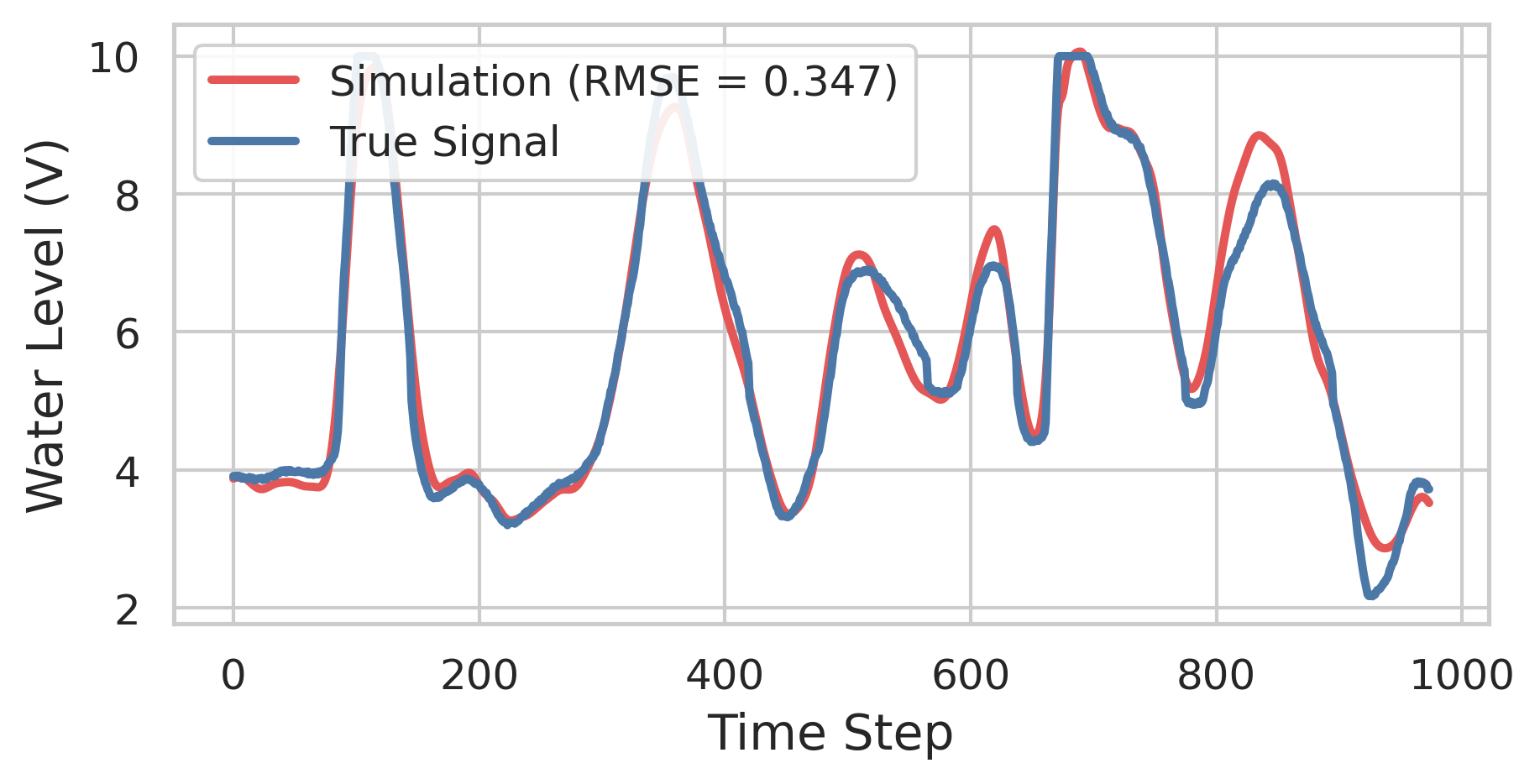}
    \caption{Simulation results on the Cascaded Tanks benchmark. The measured output is compared with the output predicted by the identified TNKM model.}
    \label{fig:tanks_test}
\end{center}
\end{figure}

The results in Table~\ref{table:tanks} indicate that TNKM achieves
predictive performance comparable with the considered baseline methods.
Among the black-box approaches, TNKM provides competitive accuracy while
requiring substantially lower training time. Compared with gray-box
methods such as SSNN and WP-LFR, which incorporate additional information
about the underlying system, TNKM achieves similar predictive performance
using a fully data-driven formulation. The recursive simulation results
in Figure~\ref{fig:tanks_test} further demonstrate that TNKM can capture
the long-horizon nonlinear behavior of the system despite the presence of
strong nonlinearities and saturation effects.

\begin{table}
\begin{center}
\caption{Comparison of identification methods on the Cascaded Tanks benchmark. }
\label{table:tanks}
\begin{tabular}{||l||c|c|c||}
\toprule
Method & RMSE & Time & W/B \\
\midrule
Normalized Gray-box SSNN & 0.221 & 1-60 min & 8 \\
WP-LFR & 0.25 & 1-60 sec & 5 \\
CT-SUBNET & 0.306 & 1-60 min & 9 \\
SUBNET & 0.37 & 1-60 min & 10 \\
PNARX & 0.417 & 1-60 sec & 10 \\
dynoNet & 0.421 & 1-60 sec & 10 \\
GPNARX & 0.839 & 1-60 min & 10 \\
\midrule
TNKM & 0.347 & $<$ 0.1 sec & 10 \\
\bottomrule
\end{tabular}
\end{center}
\end{table}

%\begin{table} \begin{center} \end{center} \end{table}

\subsubsection{\textbf{Silverbox.}}
The Silverbox benchmark is an electronic implementation of a Duffing
oscillator. The system consists of a second-order linear time-invariant
component with a third-order nonlinear feedback element, resulting in dynamics similar to those encountered in mechanical systems with nonlinear stiffness.

The benchmark provides one training sequence and three test sequences
containing 65062, 21688, 40475, and 32000 samples, respectively. The NARX
regressor is initialized using the first 50 measured output samples of
each test sequence. The input and output lag sets are defined as
$\mathcal{I}_u = \{0,1,2,3,4,8,16,32,48\}$ and
$\mathcal{I}_y = \{1,2,3,4,8,16,32,48\}$, respectively. The selected TNKM
configuration uses a TT kernel machine (TTKM) backbone with rank $R=4$
and polynomial feature maps of degree $I_d=2$.

Table~\ref{table:silverbox} shows that TNKM achieves predictive
performance comparable with leading black-box approaches such as GPNARX
and SUBNET, while requiring only several seconds for training. In
contrast, the reported training times of more complex baseline methods
can range from minutes to longer computational times. However, fully
data-driven TNKM does not achieve the same accuracy as physics-informed
approaches such as NL-LFR, highlighting the advantage of incorporating
prior system knowledge when available.

\begin{figure}[t]
\begin{center}
    \includegraphics[width=\linewidth]{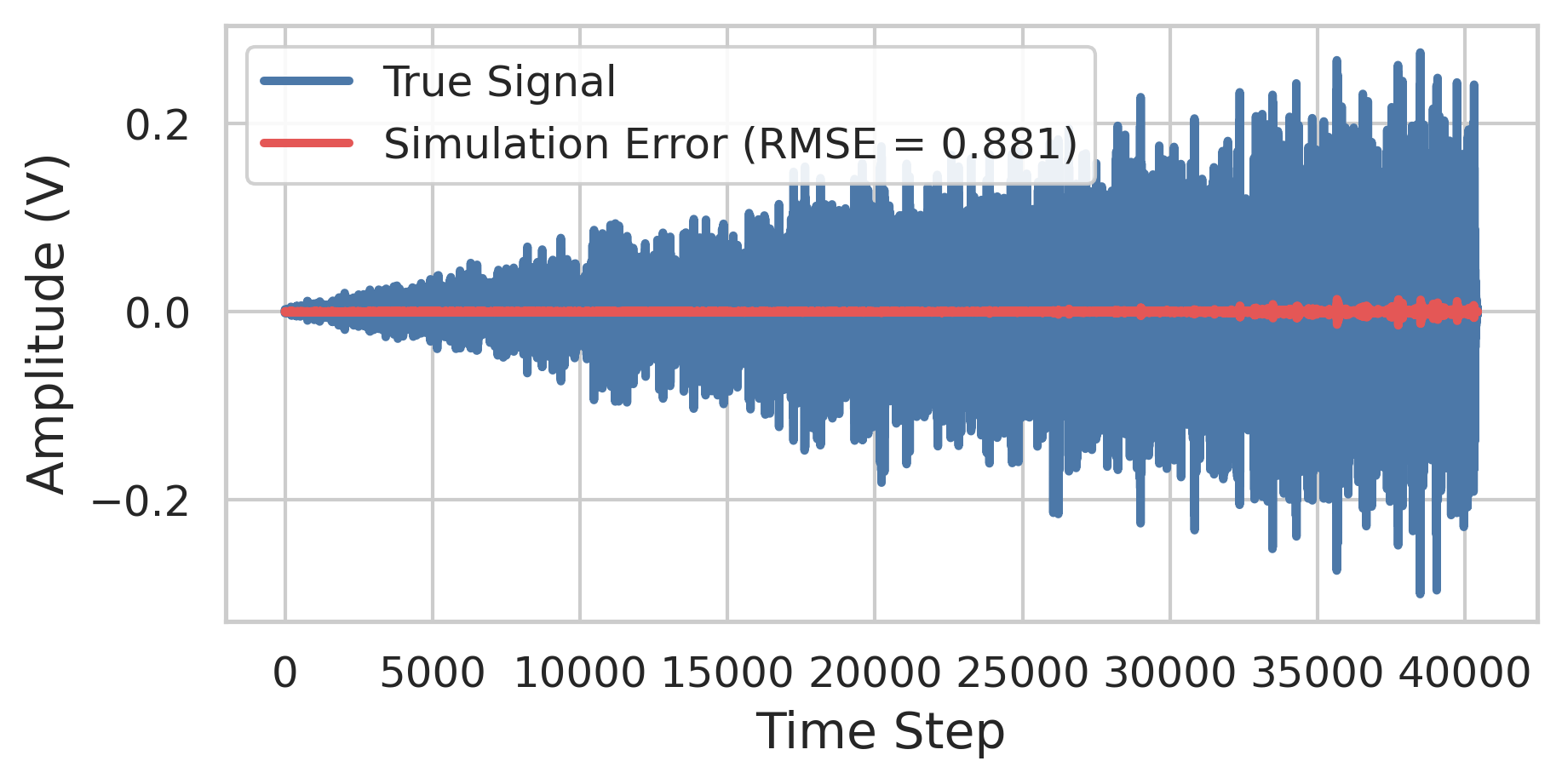}
    \caption{
    Simulation results on the Silverbox benchmark (Test 2). The figure shows the measured output together with the simulation error over the test sequence.
    }
    \label{fig:silverbox_test_2}
\end{center}
\end{figure}

Figure~\ref{fig:silverbox_test_2} shows the free-run simulation error of
TNKM on a representative test sequence. The error is shown instead of the
predicted output trajectory because the simulation accuracy is high and
the difference between the measured and predicted signals is difficult
to distinguish visually. The results demonstrate that TNKM maintains
stable long-horizon predictions over approximately 35000 simulation
steps, with only small deviations from the measured output.

This demonstrates that the proposed tensor-network parameterization can
represent nonlinear dynamical systems effectively while providing a
favorable trade-off between predictive accuracy and computational
efficiency.

\begin{table}
\begin{center}
\caption{Comparison of identification methods on the Silverbox benchmark. RMSE values are reported as RMSE $\times 10^3$.}
\label{table:silverbox}
\scalebox{0.88}{
\begin{tabular}{||l||c|c|c|c|c||}
\toprule
Method & RMSE(1) & RMSE(2) & RMSE(3) & Time & W/B \\
\midrule
NL-LFR & 0.289 & 0.334 & 0.257 & 1-60 sec & 8 \\
SUBNET & 0.36 & 1.4 & 0.32 & 1-4 days & 10 \\
GPNARX & 0.418 & 0.772 & 0.373 & 1-60 min & 10 \\
PNARX & 0.505 & 1.41 & 0.455 & 1-60 min & 10 \\
LSTM & 1.05 & 6.13 & 1.35 & 1-24 hours & 10 \\
dynoNet & 1.218 & 5.609 & 1.348 & 1-60 min & 10 \\
\midrule
TNKM & 0.402 & 0.881 & 0.368 & 1-60 sec & 10 \\
\bottomrule
\end{tabular}
}
\end{center}
\end{table}

%\begin{table} \begin{center} \end{center} \end{table}
%%\scalebox{0.9}{} %tabular

\section{Discussion}\label{sec:discussion}
\subsection{Practical implications}
The principal contribution of the \emph{tnkm} library is the unification
of the main components required for constructing tensor network kernel
machines within a single framework. By decoupling feature maps,
tensor-network parameterizations, and optimization algorithms, the
library enables systematic exploration of different TNKM configurations
while maintaining a consistent programming interface.

The library is also designed for extensibility. New features, tensor-networks, and optimization algorithms can be integrated through the existing interfaces without modifying the underlying workflow. This allows \emph{tnkm} to evolve beyond its current implementations and provides a foundation for developing new tensor-network-based learning methods.

\subsection{Relation to existing approaches}
TNKMs occupy an intermediate position between
classical kernel methods~\citep{kernel_2001_smola} and deep neural
networks~\citep{dl_2016_goodfellow}. Similar to kernel methods, TNKM
constructs nonlinear models using explicit feature representations, but
replaces the full parameter vector with a compact low-rank
tensor-network parameterization.
Compared with neural networks, TNKM relies on predefined feature maps
rather than learning representations directly from data. This provides
greater control over the model representation and facilitates the
incorporation of prior knowledge. However, the model performance becomes directly dependent on the choice of a feature map.

\subsection{Current limitations}
As with many machine learning models, the performance of TNKM depends on
the choice of hyperparameters~\citep{ho_2021_bischl}, including the
feature map $\vect{\phi}$, embedding dimensions $I_d$, tensor-network architecture $\vect{w}(\vect{v})$, tensor ranks $R$, and regularization parameters. In particular, the feature representation has a substantial impact on model performance. Unlike neural networks, where feature extraction and parameter learning are performed jointly~\citep{cnn_2015_nash}, TNKM relies on predefined feature maps. Consequently, selecting appropriate features remains an important part of the modeling process, and automatic model selection is still an open research problem.
The supported optimization methods also reflect a trade-off between
efficiency and flexibility. ALS is highly efficient for least-squares
objectives, whereas gradient-based methods support a broader range of
learning problems at the expense of additional hyperparameter tuning.

\subsection{Future perspectives}
Several directions could further extend the capabilities of the
\emph{tnkm} library. On the tensor-network side, additional
parameterizations, including tensor ring~\citep{tr_2016_zhao},
Tucker~\citep{TD_Kolda_2009}, and hierarchical
Tucker~\citep{ht_2010_grasedyck} decompositions, could offer alternative
trade-offs between compactness and expressiveness.

Another promising direction is automatic feature learning (see, e.g.,
\citep{fltnkm_2025_saiapin}), which could reduce the reliance on manually
designed embeddings while preserving the advantages of explicit feature
representations.

Probabilistic extensions of TNKM constitute another natural avenue for
future research. Bayesian inference methods~\citep{la_tnkm_saiapin_2026}, uncertainty quantification, and probabilistic priors over tensor-network parameters could enable automatic rank selection~\citep{btn_2025_kilic}, provide predictive uncertainty estimates, and establish closer connections with Gaussian process models~\citep{gp_2006_rasmussen}.

Finally, extending the collection of benchmark problems and application
examples would further demonstrate the applicability of TNKM models
across diverse domains.

\section{Conclusion}\label{sec:conclude}
This work presented \emph{tnkm}, an open-source Python library for
constructing and training tensor network kernel machines. The library
provides a unified implementation of the key components required for TNKM
modeling, including nonlinear feature representations,
tensor-network parameterizations, and optimization algorithms. By
combining these components within a common framework, \emph{tnkm}
facilitates the development and application of tensor-network-based
models for nonlinear learning problems. Experimental results on nonlinear
benchmarks demonstrate that the implemented TNKM methods achieve
competitive predictive performance while maintaining low computational costs.

\section*{Code Availability}
The source code of the \emph{tnkm} library is publicly available on
GitHub under the MIT license
\footnote{\url{https://github.com/AlbMLpy/tnkm/}}. The repository
provides installation instructions, documentation, and examples
illustrating the use of TNKM models.

\begin{ack}
This publication is part of the project Sustainable learning for Artificial Intelligence from noisy large-scale data (with project number VI.Vidi.213.017) which is financed by the Dutch Research Council (NWO).
\end{ack}

%\section*{DECLARATION OF GENERATIVE AI AND AI-ASSISTED TECHNOLOGIES IN THE WRITING PROCESS}
%During the preparation of this work the author(s) used [NAME TOOL / SERVICE] in order to [REASON]. After using this tool/service, the author(s) reviewed and edited the content as needed and take(s) full responsibility for the content of the publication.

\bibliography{references}

%\appendix
%\section{A summary of Latin grammar}    % Each appendix must have a short title.
%\section{Some Latin vocabulary}              % Sections and subsections are supported

\end{document}